\documentclass[12pt]{revtex4}
\usepackage{rotating}
\usepackage{graphicx}
\usepackage[english]{babel}
\usepackage{times}
\usepackage{mathptm}
\usepackage{supertabular}
\usepackage{dcolumn}
\usepackage{fancyhdr}

\renewcommand{\baselinestretch}{1.5}
\begin{document}
\renewcommand{\baselinestretch}{1.5}
\begin{titlepage}
\title{A DIFFERENTIAL EQUATION  FOR THE TRANSITION PROBABILITY  B(E2)$\uparrow$ 
   AND THE 
RESULTING RECURSION RELATIONS CONNECTING  EVEN-EVEN NUCLEI$*$} 

\author{{ S. Pattnaik} \\
{ Taratarini  College, Purusottampur, Ganjam, Odisha, India.     }\\
{      R. C. Nayak  } \\
{Department of Physics, Berhampur University, Brahmapur-760007,India.} }


\renewcommand{\baselinestretch}{1.5}
\begin{abstract}
  We obtain here a new relation for the reduced electric quadrupole 
transition probability
 $B(E2)\uparrow$ of
a given nucleus in terms
of  its derivatives with respect to neutron and proton numbers based on a 
similar  local energy relation in the  
Infinite Nuclear Matter (INM)  model of Atomic 
Nuclei, which is  essentially  built on the foundation 
 of the Hugenholtz-Van Hove Theorem
of many-body theory. Obviously such a   relation in the form of a differential
equation is expected  to be more powerful than the  usual 
algebraic  difference  equations. Although the relation for $B(E2)\uparrow$ 
has been  perceived simply on the basis of a corresponding differential
equation for  the local energy in  the 
INM model, its theoretical foundation otherwise has been clearly demonstrated.
 We further
 exploit  the differential equation  in using 
the very definitions of the  derivatives to obtain
two different recursion relations for    $B(E2)\uparrow$,
connecting  in each case  three neighboring even-even nuclei from lower to
 higher mass 
numbers and vice-verse. We  demonstrate their numerical validity using available
data throughout the nuclear chart  and also explore their
possible  utility in predicting $B(E2)\uparrow$ values.

\end{abstract}
\maketitle
\vskip 5cm
$*$ This is a slightly modified version of the article just published in Int.
 Jou. of Mod. Phys. {\bf E23}(2014)1450022.

\end{titlepage}
\newpage
\section{Introduction}
From the very advent of nuclear physics, ground-state properties of nuclei 
mainly mass and binding energies  have occupied special fancy of the physicists
 after the formulation of the celebrated Bethe-Weizsacker Semi-Empirical Mass
 formula, which   of course is  based on the classical liquid drop
 picture of the nucleus. In due course,
the interest was  confined to explore   possible existence of 
symmetry in 
 nuclear dynamics, in developing  mass formulas such as the 
Garvey-Kelson \cite{gkl} mass
formula that connects masses of six neighboring nuclei. Therefore it was
 natural to
 expect such symmetry to be 
 manifested in the properties other than the ground-states and  specifically
in the excited states. In this regard Ross and Bhaduri \cite{rbh}
 have  succeeded to some extent to
 find difference equations involving  the  reduced transition 
probabilities $B(E2)\uparrow$ of the neighboring even-even nuclei. 
Patnaik et al. \cite{pat}
 on the other hand have  also succeeded   in establishing
even more  simpler difference equations connecting  these values of 
four neighboring even-even nuclei. 

 Here in the present work we have made an attempt first to obtain a possible
relation involving    the  reduced transition probability 
of a given nucleus with its derivatives 
with respect to neutron and proton numbers, 
  instead of the usual difference equations. We could achieve  this  based 
on the local energy
 $\eta$-differential equation of the Infinite Nuclear Matter
(INM) model \cite{inm,in2,in3,in5} of atomic nuclei.
 In passing one may  note, that the $\eta$-relation connecting the partial
 derivatives of  the local energies in the INM model 
   is primarily responsible for the success \cite{in5} of
 the INM 
model as a mass-formula. Such a differential equation in $\eta$ was possible
in the model due to its very foundation, i. e.,    the  Hugenholtz-Van Hove(HVH) theorem \cite{hvh,hvg} of
many-body theory.  Apart from this, the INM model  has succeeded in
 resolving \cite{rcn3,in4}
 the so-called $r_0$-Paradox and also in determining the value 
of nuclear incompressibility from known masses of nuclei. 
It is needless to mention here that  
 any relation in the form of a differential equation of any physical 
quantity is intrinsically sound enough to posses a good predictive ability. This
philosophy has been well demonstrated in the INM model,  specifically for
the prediction \cite{in5} of masses of nuclei
throughout the nuclear chart.    

The local energy of a nucleus as per the INM model  basically constitutes  the
 shell and 
deformation energies,  and has been shown \cite{rcn} to carry
 the shell
structure of a given nucleus [for a comprehensive account of all these aspects
readers may refer our recently published book \cite{lrc}]. Therefore
  it is likely that this physical quantity
 $\eta$ has a good one-to-one correspondence  with the properties of 
excited states of a given nucleus in general,  and in particular 
 the reduced transition probability. 
In view of all these facts,  the local energy relation has been 
taken as the basis for formulating a similar differential equation  
connecting the  $B(E2)\uparrow$ value  of a given nucleus with its 
derivatives  with respect to neutron  and proton numbers.   In Sec.
 II, we show how such a   relation in the form of a differential equation
 can be formulated 
 followed by its possible theoretical justification.  Sec. III deals
with how the same  differential  equation can be used to derive two
recursion relations in $B(E2)\uparrow$, connecting in each case three different
 neighboring even-even nuclei.
 Finally we present in 
 Section IV, their  numerical validity   when subjected to
known \cite{rmn} experimental data through out the nuclear chart, and their 
possible utility for purpose of  predictions.

\section{Derivation of the Differential Equation involving  B(E2)$\uparrow$}

 It may be mentioned here  that one of the basic physical quantities of a given 
nucleus as per the INM model is the
the local energy $\eta$ that satisfies the INM equation \cite{in5}
\begin{equation}
  \label{eta} {\eta(N,Z)/A}={1\over 2} \left[(1+\beta){(\partial \eta
  /\partial N)}_Z + (1-\beta){(\partial \eta / \partial Z})_N\right],
\end{equation}
where Z, N, A  refer to  proton, neutron and   mass numbers 
respectively, while $\beta$ is the  usual asymmetry parameter given by 
(N-Z)/A.   $\eta$ in conjunction with
two other  physical quantities
$E(N,Z) $ and $ f(N,Z) $   defines the ground-state energy $E^F$ of
a given nucleus  given by
\begin{equation}
\label{efe}
E^{\rm\, F}(N,Z) = E (N,Z)  + f(N,Z) + \eta (N,Z).
\end{equation}
 $E$ being the energy of asymmetric nuclear matter and $f$
 being the energy of the INM sphere are global in nature, while 
$\eta$ characterizing a finite nucleus is considered local in contrast.
Consequently these three quantities are considered distinct in the sense that
each of them refers to a specific characteristic of the nucleus and as such, 
are more or less independent of each other. 
Conceptually the local energy equation (\ref{eta}) could be obtained when
Eq. (\ref{efe}) is  subjected to 
the generalized  HVH theorem \cite{hvg} of
 many-body theory given by
\begin{equation}
\label{hvh}
 {E/ A} =  [ (1+\beta)\epsilon_{\rm n}+(1-\beta)\epsilon_{\rm p} ]/2.
\end{equation}
 $\epsilon_{\rm n}=(\partial E/\partial N)_Z$ and $\epsilon_{\rm p} =
(\partial E/\partial Z)_N$ are respectively the neutron and proton Fermi energies
 of nuclear matter.

Physically the local energy $\eta$ embodies all the characteristic 
properties of a given nucleus, mainly the shell and deformation, and has been
explicitly shown \cite{rcn} to carry the shell-structure. Therefore  it is 
 likely  to have some characteristic correspondence   with  the 
properties  of excited states of a given
nucleus in general and in particular,  the reduced transition probability $B(E2)\uparrow$.
Accordingly 
 the above $\eta$-equation (\ref{eta}) can be used as an ansatz to satisfy a
similar relation involving the $B(E2)\uparrow$ of a given nucleus. As a result  we write on analogy, 
a similar  equation for $B(E2)\uparrow$ as
\begin{equation}
  \label{be2} {B(E2)[N,Z]/ A}={1\over 2} \left[(1+\beta){\Bigl(\partial B(E2)
  / \partial N\Bigr)}_Z + (1-\beta){\Bigl(\partial B(E2)/  \partial Z}\Bigr)_N\right].
\end{equation}
Thus we see that we have a relation (\ref{be2}) that  connects the $B(E2)\uparrow$
 value of a given nucleus (N,Z)
with its partial derivatives with respect to neutron   and proton 
numbers N and  Z. It is true that our   proposition of  this differential
 equation for $B(E2)\uparrow$ is 
 purely   on the basis of 
intuition and on analogy with  a similar relation for the local
energy of a nucleus in the INM model.  However  to what extent it is  true,
 that needs to be 
established. This we show in the following. 

\begin{figure}[bth]
\includegraphics[width=6.3in, height=7.0in,angle=-0]{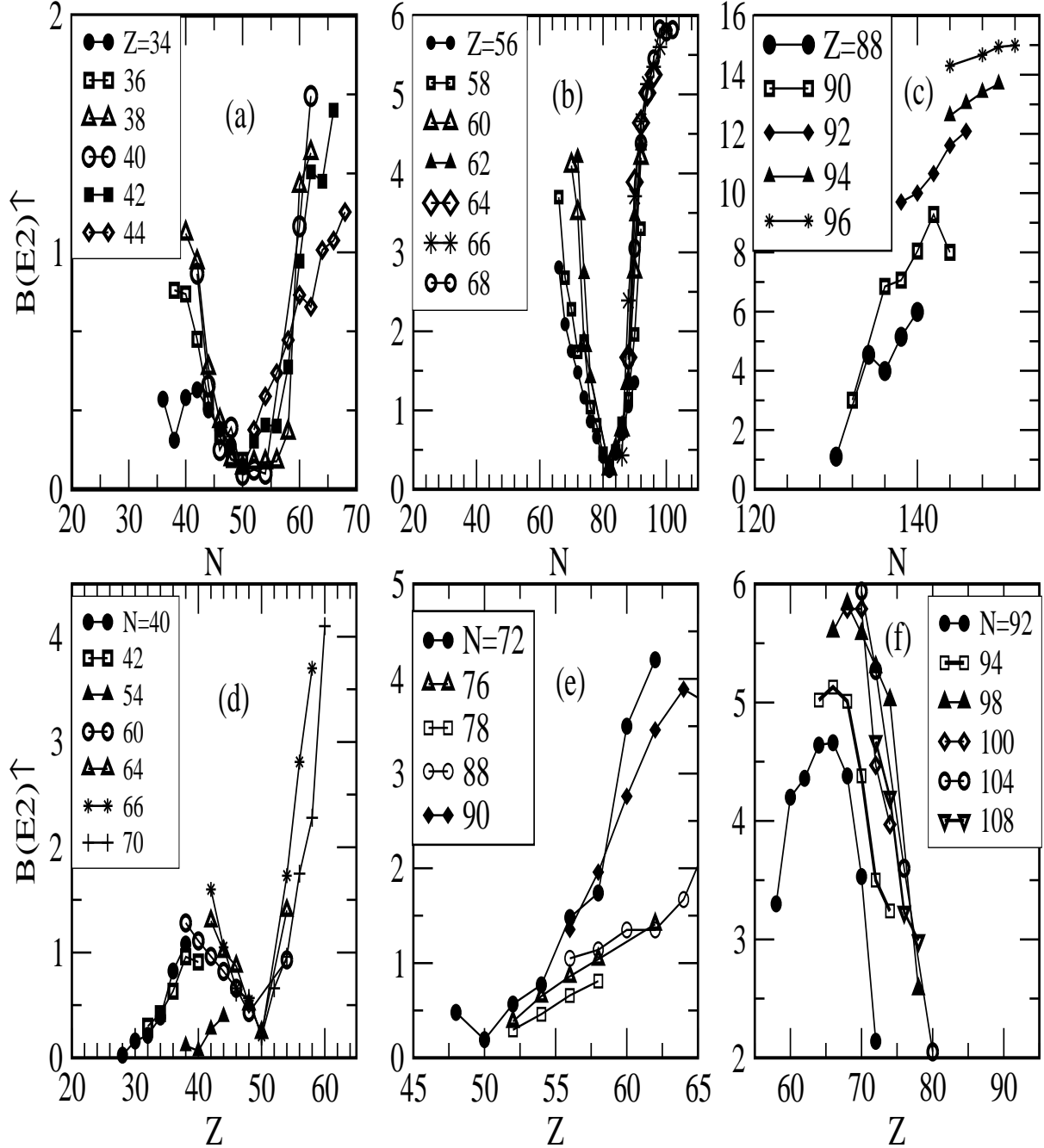}
\renewcommand{\baselinestretch}{1.}
\caption{Known $B(E2)\uparrow$ values in standard units $e^2b^2$ plotted as isolines for even-even nuclei.
 Isolines drawn  in the graphs (a-c)  connect these  values
of various  isotopes for Z=34 onwards with varying neutron number N,
 while the isolines drawn  in the graphs (d-f) show the same for 
isotones for N=40 onwards with varying proton number Z. Other possible  
isolines are not
 shown here  to avoid clumsiness of the graphs.}
\end{figure}

For a theoretical justification of the above equation, we use the
 approximation  of expressing  $B(E2)\uparrow$  as the sum of two 
different functions $ B_1(N)$ and  $ B_2(Z)$ as
\begin{equation}
  \label{f} {B(E2)[N,Z]}= B_1(N) + B_2(Z).
\end{equation}
The goodness of this simplistic approximation can only be judged from numerical
analysis of the resulting equations that follow using the experimental data. 
 Secondly we use the empirical fact [see Fig. 1] that  $B(E2)\uparrow$s are
 more or less slowly varying functions of N 
and Z  locally. This assumption however cannot be strictly true
  at magic 
numbers and in regions where  deformations drastically change. In fact known
  $B(E2)\uparrow$ values plotted as isolines for  isotopes and isotones 
 in Fig. 1, convincingly demonstrate this aspect in most of the 
cases.  The usual typical bending  and kinks at magic numbers like 50, 82 and
also across  Z$\simeq$66 [Fig. 1(f)] can be seen as a result of  sharply 
changing deformations.  Consequently
 $B_1$ and $B_2$  can be written directly
proportional to N and Z respectively as 
\begin{equation}
\label{f2}
  B_1(N)= \lambda N  ,
~~and~~   B_2(Z) =\nu Z,
\end{equation}
where $\lambda$  and $\nu$ are  arbitrary constants and vary from branch to
branch across the kinks. Then one can
easily see that  just by substitution of the above two  Eqs. (\ref{f},\ref{f2}),
 the differential Eq. (\ref{be2}) gets directly satisfied.
 Thus  the proposed differential  equation for
$B(E2)\uparrow$ analogous to the local energy relation in the INM model gets 
theoretically justified. 
 However, the  differential Eq.
 (\ref{be2}) has its own limitations, and need  not be expected to remain 
strictly valid across the   magic-number nuclei as well as in regions
where deformations sharply  change because of the very approximations involved
in proving it.

\section{Derivation of the Recursion Relations in    B(E2)$\uparrow$}

In order to utilize the differential  Eq. (\ref{be2}) for all practical purposes,
it is desirable to obtain possible  recursion relations in  $B(E2)\uparrow$ 
for even-even nuclei in (N,Z) space.
The partial derivatives occurring in this equation  at mathematical
 level are defined for continuous
 functions. However for finite nuclei, these derivatives  are to be
evaluated taking the difference of $B(E2)\uparrow$ values
of neighboring nuclei. Since our interest is to obtain recursion relations for
even-even nuclei,
  we use in the above equation the usual 
forward and backward definitions for the  partial  derivatives   given by
\begin{eqnarray}
  \label{der} 
\Bigl({\partial B(E2)/
  \partial N}\Bigr)_Z &\simeq&{1\over 2}  \Bigl[ B(E2)[N+2,Z]-B(E2)
  [N,Z]\Bigr], \nonumber \\ 
\Bigl( {\partial
  B(E2)/ \partial Z}\Bigr)_N &\simeq&{1\over 2}  \Bigl[ B(E2)
  [N,Z+2]-B(E2)[N,Z] \Bigr] ,\\
& & and \nonumber \\
\Bigl({\partial B(E2)/
  \partial N}\Bigr)_Z &\simeq&{1\over 2}  \Bigl[ B(E2)[N,Z]-B(E2)
  [N-2,Z]\Bigr], \nonumber \\ 
\Bigl( {\partial
  B(E2)/ \partial Z}\Bigr)_N &\simeq&{1\over 2}  \Bigl[ B(E2)
  [N,Z]-B(E2)[N,Z-2] \Bigr] .
\end{eqnarray}
 Substituting the above two pairs of   definitions for the derivatives in 
Eq. (\ref{be2}) separately, 
we arrive at the following two recursion 
relations for $B(E2)\uparrow$ connecting  neighboring even-even nuclei.
These are
\begin{eqnarray}
\label{b2f}
B(E2)[N,Z] &=& {N \over {A-2}}\: B(E2) [N-2,Z] + {Z \over {A-2}}\:B(E2) 
[N,Z-2]  ,\\
\label{b2b}
 B(E2) [N,Z] &=& {N \over {A+2}}\: B(E2) [N+2,Z]+{Z \over {A+2}}\: B(E2)
[N,Z+2] ..
\end {eqnarray}
The first recursion  relation (\ref{b2f}) connects three neighboring nuclei (N,Z), (N-2,Z)
and (N,Z-2) while the second one (\ref{b2b}) connects (N,Z), (N,Z+2) and (N+2,Z). Thus both
these  relations connect only three neighboring even-even nuclei.
  The first one  relates  $B(E2)\uparrow$s  
of lower  to higher mass nuclei while the second one 
relates higher to lower mass and hence they can be termed as forward and 
backward recursion relations  termed  as,
 B(E2)-F and B(E2)-B 
respectively. Thus   depending on the availability of $B(E2)\uparrow$ data, one can use 
either or both  of these two relations to obtain the corresponding
unknown  values of   neighboring nuclei.

\section{ Numerical  Test of the  Recursion Relations in B(E2)$\uparrow$}

\begin{figure}[b!t!h!]
\includegraphics[width=5.in, height=5in,angle=-0]{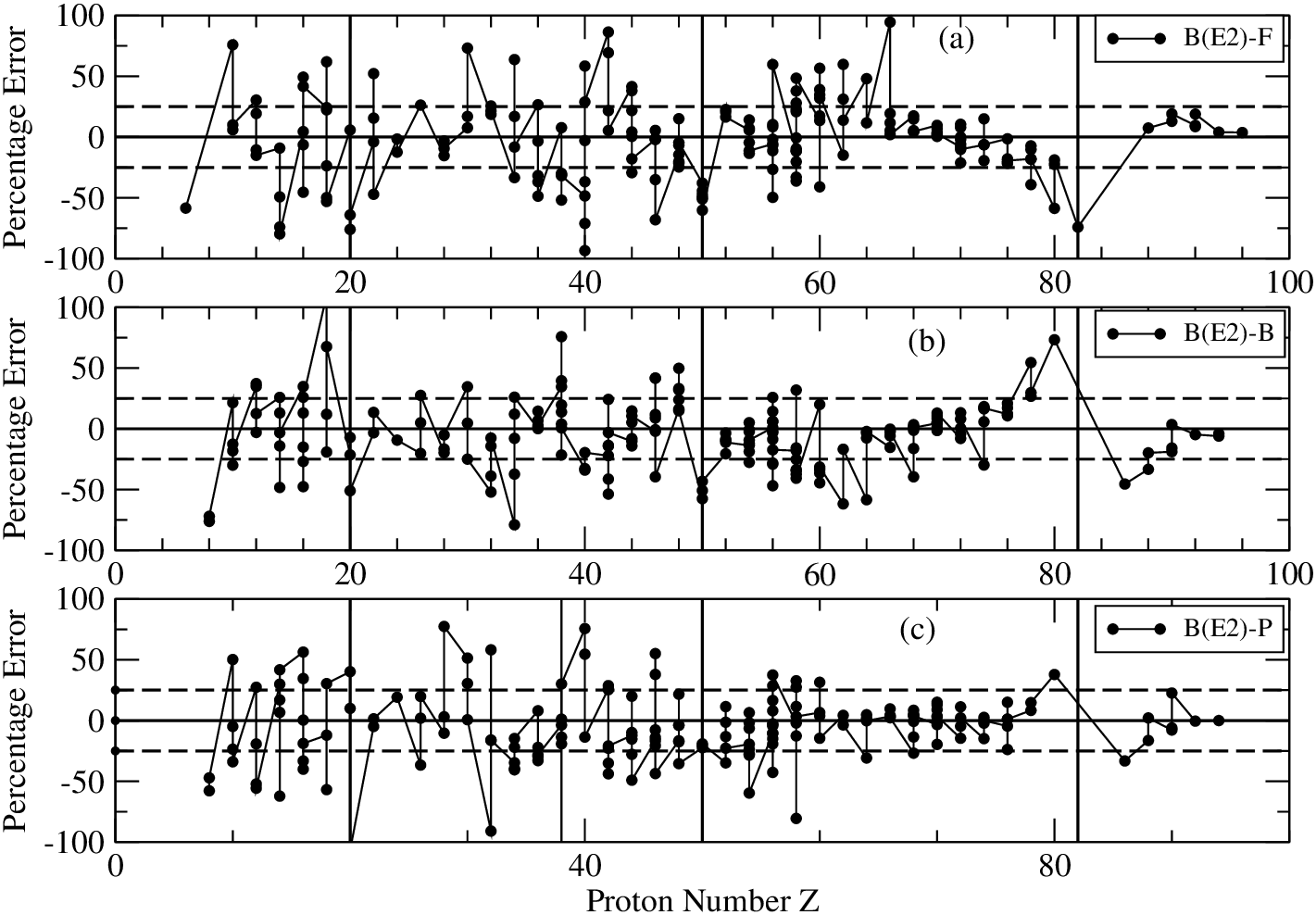}
\renewcommand{\baselinestretch}{1.}
\caption{Numerical Test of  the   Recursion Relations connecting $B(E2)\uparrow$ values 
of 
 neighboring nuclei. 
 The  percentage errors of the computed $B(E2)\uparrow$ values of all the
anchor nuclei  are plotted
 against  Proton  Number Z of those nuclei.
The graph (a)   shown as  B(E2)-F 
corresponds to the results of the relation (\ref{b2f}) while graph (b)
marked as B(E2)-B shows  those of the relation (\ref{b2b}). 
  Graph (c) marked as  B(E2)-P
presents  those of Patnaik et al. \cite{pat} [see Eq. (\ref{b2p})]. The vertical 
solid lines are drawn   just  to focus  larger deviations if any
 at the magic and semi-magic numbers. }
\end{figure}
\begin{figure}[b!t!h!]
\includegraphics[width=5.in, height=5in,angle=-0]{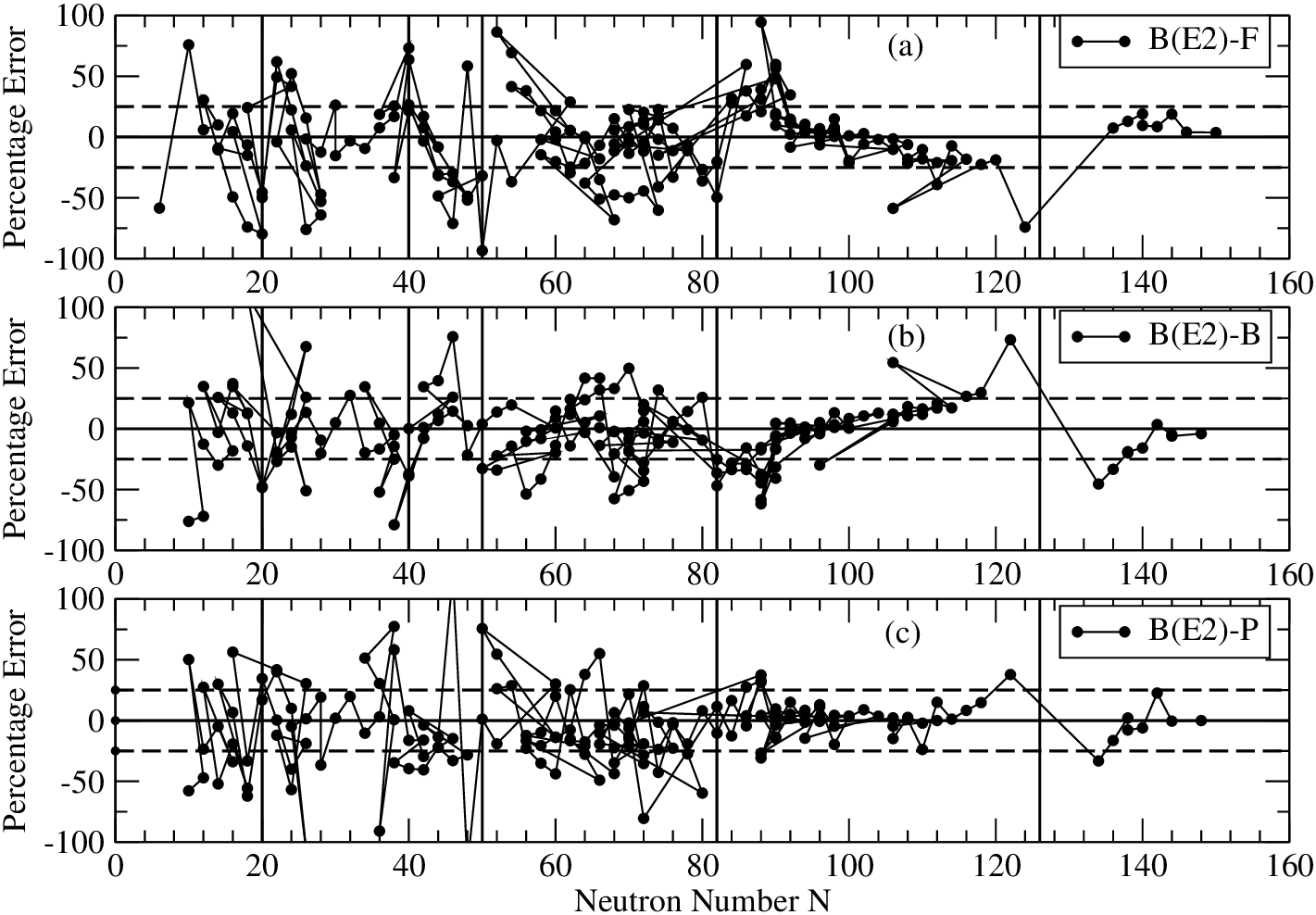}
\renewcommand{\baselinestretch}{1.}
\caption{Same as Fig.2 plotted versus Neutron Number N.}
\end{figure}
Having derived the recursion relations in $B(E2)\uparrow$  from the 
 differential equation (\ref{der}), it is desirable  to test their numerical
validity to see to what extent they 
stand the known experimental data throughout the nuclear chart. This would
also numerically support the differential Eq. (\ref{be2}) from which
 the recursion 
relations are derived.  For this purpose we use 
the experimentally adopted  $B(E2)\uparrow$ data set of   Raman et al. \cite{rmn} 
in the above  relations and
compute the same  of all possible   anchor nuclei, that are
 characterized by the neutron and proton numbers (N,Z) occurring in the left
  hand sides of the  
relations (\ref{b2f},\ref{b2b}) in the mass range of A=10 to 240.
For better visualization of our results, we calculate  the deviation of the 
computed    $B(E2)\uparrow$ values   from  those of  the experimental  data 
     in terms of the   percentage errors 
  following Raman et al. \cite{rm2}. 
The percentage error of a particular calculated quantity is  as usual
defined as the  deviation of that quantity  from
  that of the experiment
   divided by the average of
 the concerned data inputs and then expressed as the percentage of the average.
Obviously the larger the percentage error  larger is the deviation of the 
concerned computed value.
  These percentages so computed are plotted
 in the figures  2 and
3 against proton and neutron number respectively. This is done to ascertain
to what extent   large deviations occur   at proton and neutron magic
numbers.  
From the  presented results we see,  that in most of the cases both forward
and backward recursion relations  (\ref{b2f},\ref{b2b}) 
 give  reasonably good agreement with experiment.
 The deviations in  119 out of 187 cases for  
 the forward relation [B(E2)-F]  and 124 out of 186 cases for the backward
relation  [B(E2)-B] lie within 
$\pm$25\% error [shown   within broken lines in the figures]. 
 It should be noted  that even the
 experimental
data themselves vary  quite widely from one experiment to another for the same
nucleus [ see for 
instance Table II of the data compilation by Raman et al. \cite{rmn}]. 
Just to cite two examples of such wide variation, the $B(E2)\uparrow$ value for the nucleus
 $ ^{142}Nd$ lies  in the
 range  0.256  to 0.437 units, while the adopted value is 0.265 units
and the same for $^{154}Sm$  varies from 3.45 to 6.8 units, the adopted value
being  4.36 units \cite{rmn}. Such wide variation in the experimental values 
are almost prevalent in most of the cases. 
  In view of this, the agreement of the model recursion relations with those
of experiment can be considered rather good. 
 However one can see from the figures 2 and 3, that  the percentage errors (deviations) are relatively
higher for some nuclei in the neighborhood of the  magic numbers 20, 82, 126
and semi-magic number 40.  Such increase in the vicinity of the magic 
numbers is  expected as the differential Eq. (\ref{be2}) from which the recursion
relations are derived, need not be  strictly valid
 at the magic numbers.

For sake of comparison of our  results  with those of 
 algebraic relations of other authors,
 we  present here  
calculated values using the difference equations of    Patnaik et 
al. \cite{pat}. One should note that they used the sum 
rule approach of the Garvey-Kelson type  given by
\begin{equation}
\label{b2p}
B(E2)[N,Z] = B(E2)[N+2,Z]+B(E2)[N,Z+2]-B(E2)[N+2,Z+2].
\end{equation}
It may be easily noted that this difference equation connects four nuclei namely
(N,Z), (N+2,Z), (N,Z+2) and (N+2,Z+2) in contrast to three  in our 
recursion relations (\ref{b2f},\ref{b2b}). These results
are presented as B(E2)-P in the  graph (c)  of figures  2 and 3. As can be seen that
 these results are almost similar to ours. Numerically we find that the 
deviations
  up to 25\% error from experiment exists  in 117 out of 178 cases. 

\begin{figure}[b!t!h!]
\includegraphics[width=4in, height=5in,angle=-0]{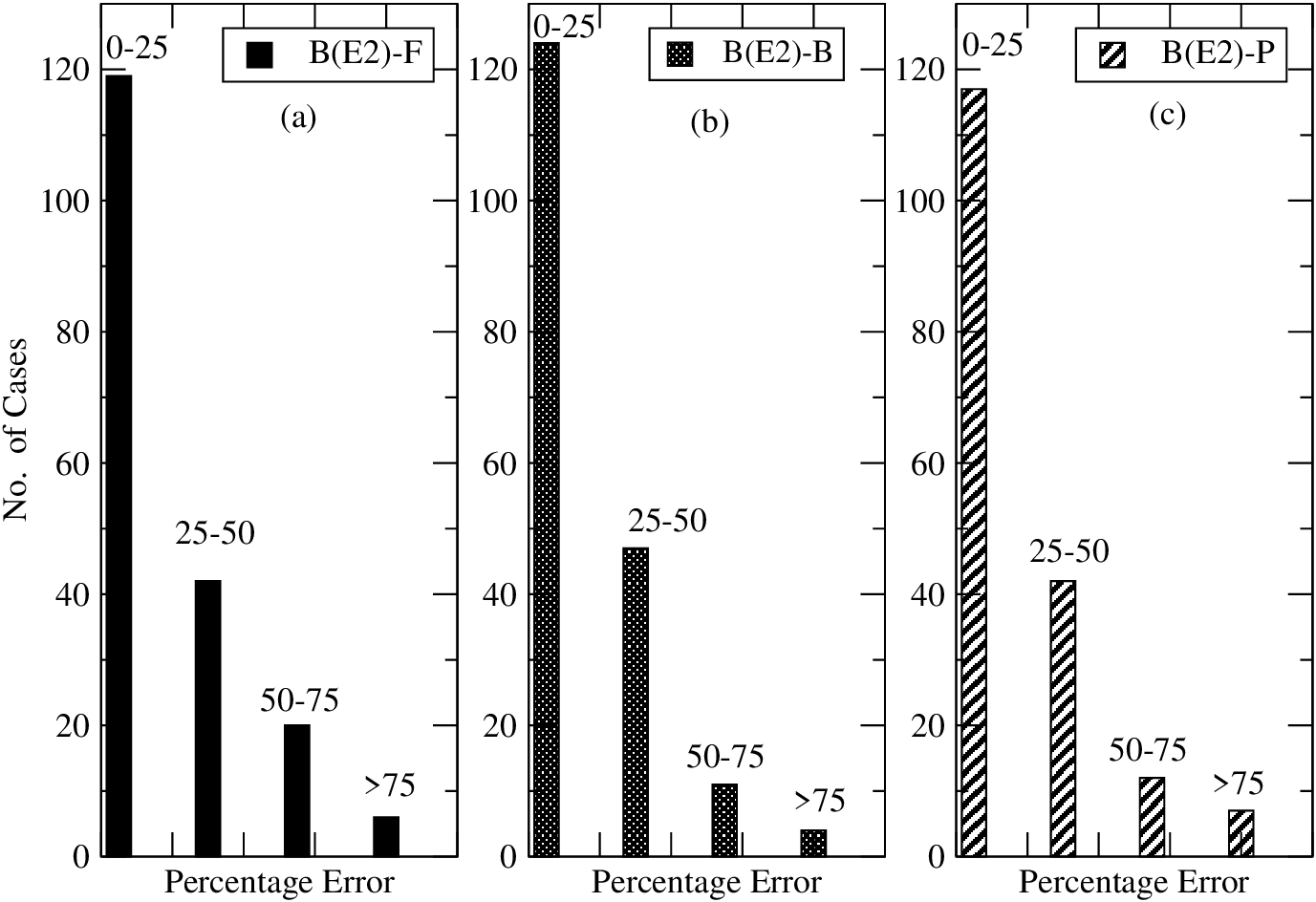}
\renewcommand{\baselinestretch}{1.}
\caption{Vertical pillars
 showing  number of cases having different ranges of absolute percentage 
 errors. Those  marked as B(E2)-F and   B(E2)-B correspond
to  results of our relations (\ref{b2f} and \ref{b2b}) respectively, whereas 
those of B(E2)-P correspond to  the Eq. (\ref{b2p}) of Patnaik et al. \cite{pat}.}
\end{figure}

 To  bring out the contrasting features of our results in a better way, 
   we also present our results   in Fig. 4 
 in the form of histograms, which represent number of cases having
different ranges of percentage errors. As can be seen,  the sharply decreasing
  heights of the vertical pillars with the increasing  range of 
 errors  are a clear testimony of the goodness  of our  recursion 
relations. Similar behavior can be   seen also with those of Patnaik et al.
\cite{pat}.

For exact numerical comparison,
   we  present in Table I,  results  obtained in our calculation
 along with those of  experiment \cite{rmn}
for some of the nuclides randomly chosen all over the  nuclear chart.
 One can easily see that the agreement of the predictions with the
 measured values is
exceedingly good in most cases. In few cases such as $^{114}Pd$, $^{124}Sn$
there exists some discrepancy in between the predicted value and those of
 experiment. 
 But we should also note, that the actual experimental data 
  for  these two nuclei
  lie  in the range of  0.203-0.380 and  0.133-0.220  units respectively, which
we  quote here from  Table II of Raman et al. \cite{rmn}.
Therefore in the light of such variations in the experimental data, predictions
of our recursion relations appear to be reasonably good and hence can be 
relied upon.

Now coming to the possible agreement in between the predictions of the
  two recursion relations (\ref{b2f}) and (\ref{b2b}) 
, we see that more or 
less they agree well. However there exists   some disagreement  in between them
in few cases such as $ ^{84}Kr, ^{148}Ce $ and $ ^{236}U$.
This may be attributed to the widely varying experimental inaccuracy of the
concerned data inputs in our recursion relations.
 For instance, in case of $^{236}U$, the predictions differ from each other
by about 17\% while  the corresponding data  inputs $^{234}U,
^{234}Th$ and $^{238}U,^{238}Pu$ have experimental error bars  
 0.20,0.70,0.20 
and 0.17 units respectively. Thus all  except the second data input
has  a relatively higher  error,
   thereby possibly causing the difference in predictions. Thus in general both
the recursion relations more or less can be relied upon for predicting the 
$B(E2)\uparrow$ values.


\begin{table}[h!t!b!]
\begin{center}
\begin{tabular}{|c c c c c c c c|}
\hline
\multicolumn{1}{|c } {Nucleus} &
\multicolumn{1}{ c }{Experiment} &
\multicolumn{1}{ c }{B(E2)-B} &
\multicolumn{1}{ c }{B(E2)-F} &
\multicolumn{1}{ c } {Nucleus} &
\multicolumn{1}{ c }{Experiment} &
\multicolumn{1}{ c }{B(E2)-B} &
\multicolumn{1}{ c|}{B(E2)-F} \\
\hline
$ ^{26}Ne$ &0.0228 (41) &- & 0.028 & $ ^{30}Mg$ &0.0295 (26) &0.034 & 0.026  \\
$ ^{36}Si$ &0.019 (6)  & -  &0.019   & $ ^{42}S$ &0.040 (6) & -  &0.031   \\
$ ^{46}Ar$ &0.0196 (39) & 0.035  &-  & $ ^{48}Ca$ &0.0095 (32)  & 0.019  \\
$ ^{44}Ti$ & 0.065 (16) &0.068  &-   &  $ ^{74}Ge$ & 0.300 (6)&- &0.325 \\
$ ^{80}Se$ & 0.253 (6) &- &0.196 & $ ^{84}Kr$ & 0.125 (6) &0.211  &0.122 \\
$ ^{98}Sr$ & 1.282 (39)   &-   &1.274 & $ ^{102}Zr $ &1.66 (34)   & 1.256 &-  \\
$ ^{106}Mo$ & 1.31 (7)   &-   &1.341 & $ ^{110}Ru$ & 1.05 (12) &1.269  &0.949 \\
$ ^{114}Pd$ & 0.38 (12) &0.881     &0.585 & $ ^{120}Cd$ &0.48 (6) & - &0.418  \\
$ ^{124}Sn$ &0.166 (4)   & 0.354   &-   & $ ^{128}Te$ &0.383 (6)   & - &0.432 \\
$ ^{134}Xe$ &0.34 (6) &-   &0.375       & $ ^{144}Ba$ &1.05 (6) &-   &1`.254  \\
$ ^{148}Ce$ &1.960 (18)  &1.241 &3.047 & $ ^{152}Nd$ &4.20  (28) & 3.013 &-  \\
$ ^{154}Sm$ &4.36 (5) & 3.807  &-      & $ ^{158}Gd$ &5.02 (5) &- & 5.136   \\
$ ^{162}Dy$ &5.35 (11)   &5.244  & 5.471 & $ ^{168}Er$ &5.79 (10)  &- & 5.739 \\
$ ^{174}Yb$ &5.940 (6)   &-  & 5.228 & $ ^{178}Hf$ &4.82 (6)  &5.342 & 4.450 \\
$ ^{186}W$ &3.500  (12) &-  & 2.970  & $ ^{190}Os$ &2.35 (6) &2.880 & 1.987  \\
$ ^{196}Pt$ &1.375 (16) &-    & 1.034    & $ ^{202}Hg$ &0.612 (10) &- &0.319 \\
 $ ^{220}Rn$ &1.86 (7) &-    &3.189  & $ ^{226}Ra$ &5.15 (14)   &-    &6.350 \\
 $^{232}Th$ &9.28 (10) &- &8.955 & $ ^{236}U$ &11.61 (15)   & 9.705 & 12.189 \\
 $ ^{242}Pu$ &13.40 (16)  &-    &13.949 & $ ^{246}Cm$ &14.94 (19)  &14.401 &- \\ 
\hline
\end{tabular}
\caption{Comparison of the predicted referred to as B(E2)-B [Eq. (\ref{b2b})]
and  B(E2)-F  [Eq. (\ref{b2f})] and experimentally adopted \cite{rmn}  $B(E2)\uparrow$
 values presented in 
terms of the standard units[$e^2b^2$]. The bracketed numbers refer to 
uncertainties in the last digits of the experimentally quoted values.} 
\end{center}
\end{table}
 
Once we establish the goodness of the two recursion relations, it is desirable
 to compare our predictions with
 the latest experimentally adopted data  of  Pritychenko et al. \cite{prt}. 
  It must be made
clear that none of the values of the new experimental data set has been
 used in our recursion relations. Rather  we 
use only the available data set of Raman et al. \cite{rmn} to generate all possible values of a given
nucleus employing the two recursions relations 
 (\ref{b2f}) and (\ref{b2b}). One  should  note here that each of these
relations can be rewritten in three different ways just by shifting the three
terms occurring in  them from left to right and vice-verse. 
Thus altogether,  these two relations in principle can generate up to six
 alternate
values for a given nucleus  subject to availability of the corresponding data.
Since each of the values is equally probable, 
the predicted value for a given nucleus is then obtained by the arithmetic  
mean of all those  generated values so obtained. 
  Our predictions here are confined  only for those
isotopes for which  measured values were quoted by Pritychenko et al. \cite{prt}.
The  predicted values so obtained  termed as  the 
Model values  are presented  in Table II
 for various isotopes of Z=24, 26, 28 and 30 
 along with those of the latest  experimental \cite{prt} data.

\begin{table}[h!t!b!]
\begin{center}
\begin{tabular}{|c c c c c c|}
\hline
\multicolumn{1}{|c}{ Nucleus} &
\multicolumn{1}{c}{Experiment \cite{prt}} &
\multicolumn{1}{c|}{Model} &
\multicolumn{1}{c}{ Nucleus} &
\multicolumn{1}{c}{Experiment \cite{prt}} &
\multicolumn{1}{c|}{Model} \\
\hline
$ ^{46}Cr$ & 0.093 (20) &0.1033 & $ ^{48}Cr$ & 0.137 (15)  &0.131    \\
$ ^{50}Cr$ & 0.1063 (32) &0.1107  & $ ^{52}Cr$ & 0.0627 (27) &0.0735   \\
$ ^{54}Cr$ & 0.0879 (55) &0.1030    & $ ^{56}Cr$ & 0.055 (19)&0.1109    \\
$ ^{58}Cr$ & 0.099 (28) &0.057     & $ ^{54}Fe$ & 0.0608 (31)  & 0.0750   \\
$ ^{56}Fe$ & 0.0975 (27)  & 0.0781    & $ ^{58}Fe$ & 0.123 (4)   & 0.1115   \\
$ ^{60}Fe$ & 0.0938 (88)  &0.1133   & $ ^{62}Fe$ & 0.1028 (90)  &0.081   \\
$ ^{64}Fe$ & 0.178 (17)  &0.039    & $ ^{54}Ni$ &0.061 (12)  & 0.054   \\
$ ^{56}Ni$ &0.0453 (86)  & 0.0332  & $ ^{58}Ni$ &0.0673 (17)  & 0.0751  \\
$ ^{60}Ni$ &0.0914 (17)  & 0.1110  & $ ^{62}Ni$ &0.0893 (21)  & 0.1041  \\
$ ^{64}Ni$ &0.0629  (32) & 0.0743  & $ ^{66}Ni$ &0.0611 (67)  & 0.0672  \\
$ ^{68}Ni$ &0.0260 (40)  & 0.1084  & $ ^{62}Zn$ &0.1224 (59)  &0.1591  \\
$ ^{64}Zn$ &0.1484 (52) &0.1341   & $ ^{66}Zn$ &0.1371 (29) &0.1544   \\
$ ^{68}Zn$ &0.1203 (25) &0.1626   & $ ^{70}Zn$ &0.1525 (75) &0.1399   \\
$ ^{72}Zn$ &0.174 (21)   &0.261    & $ ^{74}Zn$ &0.200 (10)   &0.207   \\
\hline
\end{tabular}
\end{center}
\renewcommand{\baselinestretch}{1.}
\caption{Comparison of the predicted  and the latest experimental \cite{prt}
  $B(E2)\uparrow$ values presented in 
terms of the standard units[$e^2b^2$]. The bracketed numbers refer to uncertainties in the last digits of the experimentally quoted values.} 
\end{table}

 The same are also plotted  in  Fig. 5 to get a  better view 
 of the results. Our predictions are confined  only for those
isotopes for which  measured values were quoted by Pritychenko et al. \cite{prt}.
 One can easily see that in all  the cases except for 
 $ ^{64}Fe $ and $^{68}Ni $, the agreement
 between the predictions
with those of the experiment are remarkably  good. For these two nuclei, the 
 discrepancies  may be attributed to the possible
 sub-shell effect as the neutron  numbers  are 38 and 40 respectively. 
For sake of comparison we have also presented in Fig. 5 results obtained
 from two shell-model calculations \cite{gxp,prt} marked here as SM1 and SM2 by
 employing two different effective interactions  GXPF1A \cite{gxp} and
 JUN45 \cite{jun}. One should note here, that the first one because of its own
limitations did  not succeed in getting reliable values for nuclei having 
neutron number beyond N=36. Hence the second shell-model with JUN45 effective 
interaction was performed by
\begin{figure}[b!t!h!]
\includegraphics[width=5.in, height=6.0in,angle=-0]{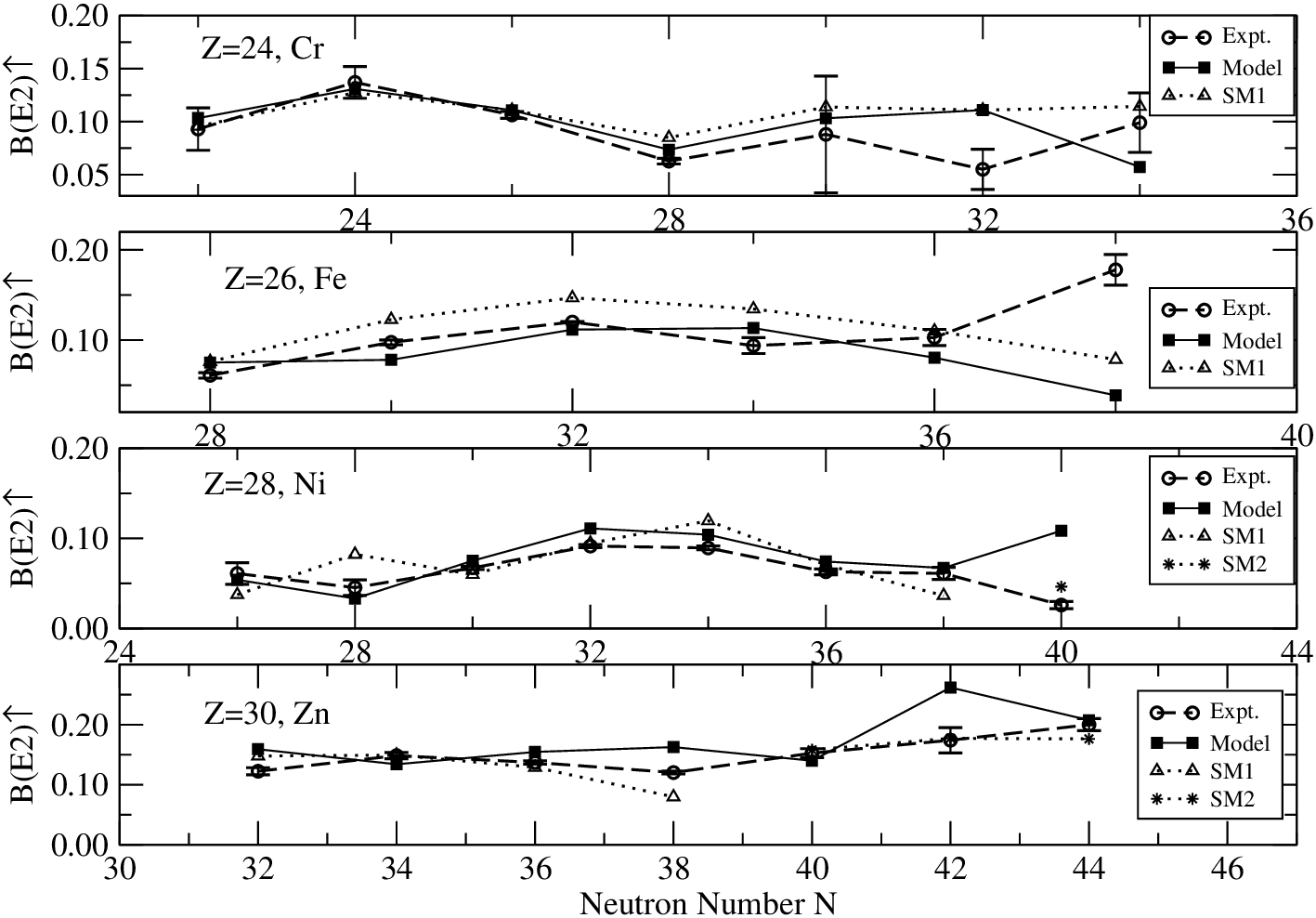}
\renewcommand{\baselinestretch}{1.}
\caption{Both calculated (Model) and the latest experimental \cite{prt} 
$B(E2)\uparrow$ values in standard units $e^2b^2$
 [see text for details]  
are presented   for various isotopes of Z=24, 26,
 28 and 30  versus 
 Neutron Numbers N. Values for different isotopes  are connected  by solid or
 dashed  lines. Experimental uncertainties are shown as vertical error bars
as usual. Recent shell-model calculated values [SM1 and SM2]  are also presented
for sake of comparison}
\end{figure}

\noindent Pritychenko et al. \cite{prt} 
for the nuclei  $ ^{64}Fe, ^{68}Ni$ and  $^{72,74}Zn$. One can 
easily see that the shell-model values SM1 almost agree with those of ours
 for almost all the isotopes. 
Even for $^{64}Fe$ where there is a little bit of discrepancy in between ours
 and the experiment, SM1 value is closer to ours. SM2 values more or less 
agree with  those of experiment.

	We have just demonstrated as shown above the utility of the recursion
 relations for predicting $B(E2)\uparrow$ values for  some of the even-even isotopes
 of Cr, Fe, Ni and Zn in agreement with the latest experimental \cite{prt} data. Therefore it is desirable to find out whether
the model is   good enough   for such predictions in the higher mass regions of
 the nuclear chart. However in these regions there is no new data to compare
with and hence we can only compare with the  data  set of Raman et al. \cite{rmn}.
With this view,  we repeated our calculations 
 for higher isotope series 
following the same methodology outlined above. 
 Since the main aim of our
 present investigation is just to  establish the goodness of our model, we
 present  
 here results of only  few such series for which experimental data exist
for a relatively large number of isotopes. Accordingly  we have chosen
four isotope  series  Z=40, 48, 66 and 78 covering 
  nuclei both in mid-mass and heavy-mass regions. Our choice of the first
two  series namely Z=40 and 48 
 is again to see to what extent our model works across the semi-magic number
40 and magic number 50.  Our predictions  along with 
 those of experimental data of     Raman et al. \cite{rmn} are presented  
in Fig. 6. From the presented results,  it is fair enough to say that the 
agreement of the model values 
  with experiment in most of the cases is  good. It is remarkable that even
sharply changing deformations at neutron numbers N=46 and 56 for Zr [Fig. 
6. (a)] are
 well reproduced.
However we find  few 
exceptions such as $^{116}Cd, ^{122}Cd,  ^{156}Dy,
$  and $^{184}Pt$,  for which  some  discrepancies exist.
Since we use the recursion relations for our predictions, such random
 cases  may be attributed to the possible
experimental inaccuracy of  $B(E2)\uparrow$ values in their
 neighborhood. For instance,
the experimental inaccuracy  in case of $^{122}Cd$ is  almost 66\% of the 
quoted value [Fig. 6 (b)], which is rather much larger than the discrepancy
of our model value. Use of such input data in the recursion relations 
are possibly  affecting  predicted values in few cases. 
we have also shown in Fig. 6 results of the two well-known theoretical  models 
namely Finite Range Droplet Model (FRDM) \cite{frd} and Single-Shell Asymptotic
Nilsson Model (SSANM) \cite{ssa} just for sake of comparison. One can see that
in most of the
cases for these two models, the discrepancies from experiment  are rather high. 

Now taking stock of all the results discussed so far,  we can fairly say that
 the
recursion relations work reasonably  well almost throughout the nuclear chart.
 Even across the
magic numbers and sharply changing deformations, these relations have succeeded
in reproducing the experimental data to a large extent with a little bit
of deviation here and there.  Elsewhere  there
exists few exceptions but the discrepancies are not that large as to make the
relations untrue. In a nutshell, the recursion relations 
for the reduced transition probability  $B(E2)\uparrow$ derived here 
can be termed   sound enough as to have passed  the numerical test both in 
reproducing and predicting  the experimental
values,  and thereby vouches for the goodness of the differential equation 
(\ref{be2}) from which they originate.
  
\begin{figure}[bth]
\includegraphics[width=6.in, height=7.0in,angle=-0]{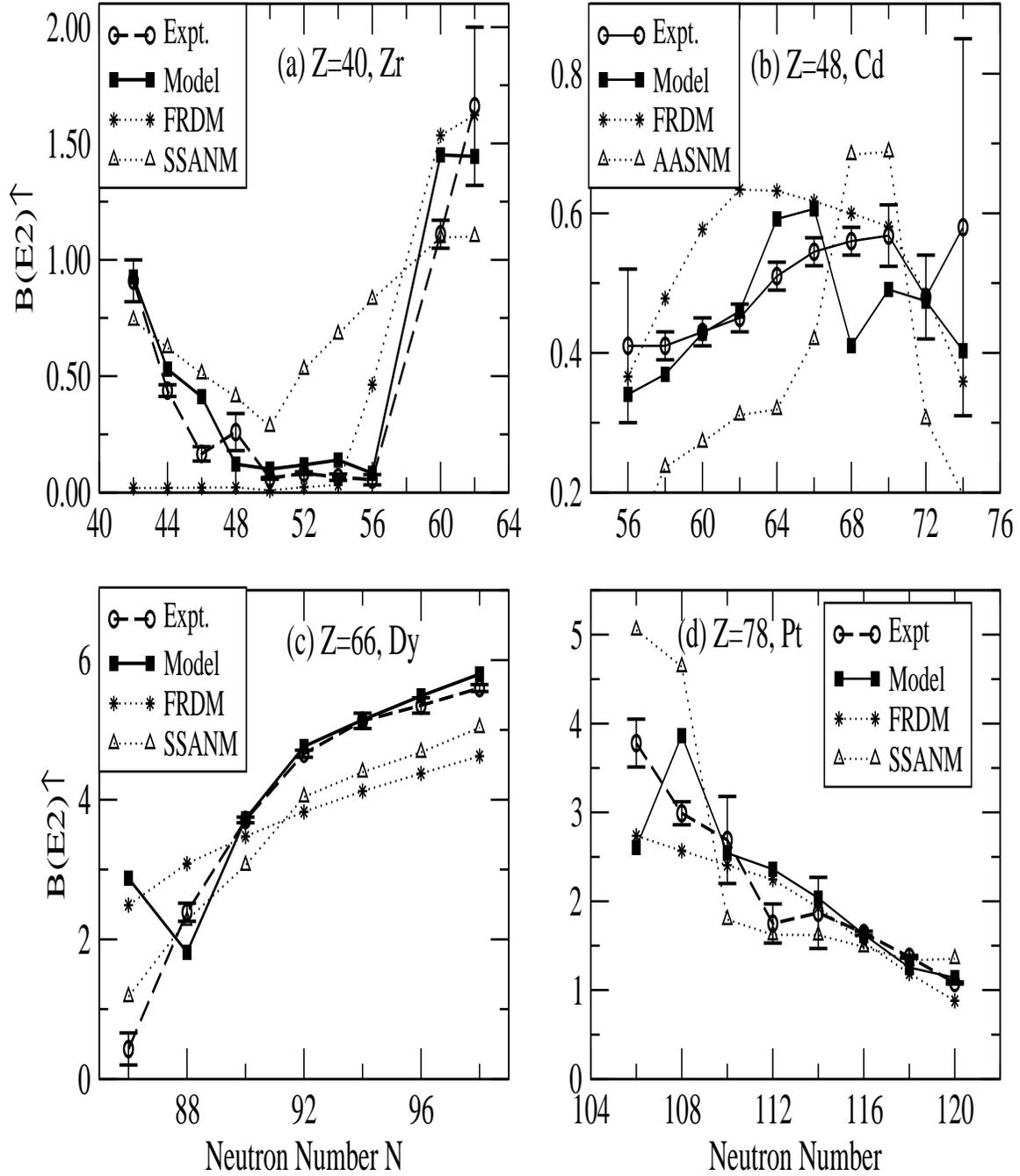}
\renewcommand{\baselinestretch}{1.}
\caption{ Similar to Fig. 5 but for Z=40, 48, 66 and 78. The experimental
data points marked as Expt
 correspond to those of Raman et al. \cite{rmn}. Data points marked as FRDM and
 SSANM correspond to  predictions of the other two theoretical models (see text).  }
\end{figure}
	\section{ Concluding Remarks}

In conclusion  we  note, that  we have  succeeded in deriving
 for the first time, a novel relation 
for the reduced transition probability $B(E2)\uparrow$ of a given nucleus
 in terms of
its derivatives with respect to neutron and proton numbers.
We could obtain such a  differential equation  on the basis of  one-to-one 
correspondence with the local energy, that   satisfies  a 
similar   relation
formulated in the INM  model of atomic nuclei. 
   We have also succeeded in 
establishing its theoretical foundation using the empirical fact, that
 $B(E2)\uparrow$s
are more or less slowly varying functions of neutron and proton numbers except
across the magic numbers.
We  further exploited the usual definitions of the two derivatives with respect
to neutron and proton numbers  occurring in
 the equation,  to derive two recursion relations in $B(E2)\uparrow$. Both these relations are
found to connect 
three neighboring even-even nuclei from lower to higher mass  and
vice-verse.
Their  numerical validity was then established using
 known experimental data set compiled by Raman et al. \cite{rmn}
 in the mass range of A=10 to 240.  
Apart from this 
 their utility was further established by comparing our predictions 
with the latest experimental data set of Pritychenko et al. \cite{prt} for the
 isotopes of Cr, Fe, Si and Ni.
The results so obtained convincingly show
the goodness of the recursion relations and thereby their parent differential
equation. 

{99}
\end{document}